\documentstyle[12pt,amsbsy]{article}
\begin{document}
\setlength{\unitlength}{1mm}
\newcommand{\te}{\theta}
\newcommand{\bee}{\begin{equation}}
\newcommand{\ene}{\end{equation}}
\newcommand{\tra}{\triangle\theta_}
\vspace*{1cm}
\begin{center}
{\Large\bf Coherent States for Particle Beams}
\end{center}
\begin{center}
{\Large\bf in the Thermal Wave Model}
\end{center}
\bigskip\bigskip

\begin{center}
{{\bf S. De Nicola}$^{1}$, {\bf R. Fedele}$^{2,3}$, {\bf V.I.
Man'ko}$^{2,4}$ and
{\bf G. Miele}$^{2,3}$}
\end{center}

\bigskip

\noindent
{\it $^{1}$ Istituto di Cibernetica del C.N.R., Arco Felice,
I--80072 Napoli, Italy}\\

\noindent
{\it $^{2}$ Dipartimento di Scienze Fisiche, Universit\`a di Napoli
''Federico II'', Mostra D'Oltremare Pad. 20, I--80125 Napoli, Italy}\\

\noindent
{\it $^{3}$ Istituto Nazionale di Fisica Nucleare, Sezione di Napoli,
Mostra D'Oltremare Pad. 20, I--80125 Napoli, Italy}\\

\noindent
{\it $^{4}$ Lebedev Physics Institute, Leninsky prospect 53, 117333
Moscow, Russia}

\bigskip\bigskip\bigskip

\begin{abstract}

In this paper, by using an analogy among {\it quantum mechanics},
{\it electromagnetic beam optics in optical fibers}, and {\it charge
particle beam dynamics},
we introduce the concept of {\it coherent states} for charged particle
beams in the framework of the {\it Thermal Wave Model} (TWM).
We give a physical meaning of the Gaussian-like
coherent structures of charged particle distribution
that are both naturally and artificially produced in an accelerating
machine in terms of the concept of coherent states widely used in
quantum mechanics and in quantum optics. According to TWM,
this can be done by using a Schr\"{o}dinger-like equation for a
complex function, the so-called {\it beam wave function} (BWF), whose
squared modulus is proportional to the transverse beam density
profile, where Planck's constant and the time are replaced by the
transverse beam
emittance and by the propagation coordinate, respectively.
The evolution of the particle beam, whose initial BWF is assumed to be
the simplest coherent state (ground-like state) associated with the beam,
in an infinite 1-D quadrupole-like device with
small sextupole and octupole aberrations, is analitically and
numerically investigated.

\end{abstract}
\baselineskip22pt

\bigskip\bigskip\bigskip


\bigskip\bigskip


\newpage

\section{Introduction}

It has been recently pointed out that the charged particle beam transport
through a generic optical device can be described by means of the so-called
{\it Thermal Wave Model} (TWM) for particle beam dynamics \cite{fm1}. According
to this approach the transverse (longitudinal) dynamics of a particle beam is
described in terms of a complex function, called {\it beam wave function}
(BWF), whose squared modulus gives the transverse (longitudinal) density
profile of the beam \cite{fm1}-\cite{fm2}, \cite{fgm}
(\cite{fpv},\cite{fmpv},\cite{fv1},\cite{fmp}). The BWF satisfies a
Schr\"{o}dinger-like equation where Planck's constant is replaced by the
transverse (longitudinal) beam emittance \cite{fm1}-\cite{fm2},\cite{fgm}
(\cite{fpv},\cite{fmpv},\cite{fv1},\cite{fmp}).

This model has been successfully applied to a number of linear and nonlinear
problems of particle beam dynamics \cite{fm1}-\cite{fmp}. In particular, it has
been used in the transverse dynamics for describing the optics of a charged
particle beam in a thin quadrupole-like lens with small sextupole and octupole
deviations \cite{fgm}, and for estimating the luminosity in linear colliders
\cite{fm2}. TWM has also been applied for describing the self-consistent
interaction of a relativistic electron-positron beams with a
collisionless, overdense plasma \cite{fs}, reproducing the main results for the
beam filamentation threshold and the self-bunching equilibrium \cite{fs}.\\TWM
seems also useful for describing longitudinal particle bunch dynamics in both
conventional accelerators and plasma-based accelerators for which coherent
instability, in terms of modulational instability, and soliton formation has
been investigated \cite{fpv},\cite{fmpv},\cite{fv1}.
Recently, TWM has been used for
describing the longitudinal dynamics of a relativistic charged particle bunch
in a circular accelerating machine, when the {\it radio-frequency} (RF)
potential well and
the self-interaction (wake fields) are present \cite{fmpv} as well as
synchrotron radiation damping and quantum excitation (photon noise) are taken
into account \cite{fmp}.

The paper is organized as follows: in section $2$ we briefly review the basic
features of the {\it Thermal Wave Model} and introduce the concept of {\it
coherent state}
for charged particle beam dynamics in analogy with the light beam optics.
In section $3$, we extend the perturbative approach previously discussed for a
{\it thin} lens \cite{fm2}, \cite{fgm}
to investigate the stationary propagation  of
a charged particle beam through an {\it infinite } 1-D
quadrupole with small {\it sextupole}
and {\it octupole} deviations. The initial BWF is assumed to be in the
ground-like state of the harmonic oscillator (pure Gaussian profile).
Numerical results for the BWF obtained with this approach are
presented and discussed. Finally, in section $4$ we give our
remarks and conclusions.

\section{The concept of coherent state in particle beams}

Let us consider a relativistic charged particle beam travelling along $z$-axis
with velocity $\beta c$ ($\beta\approx 1$). Denoting by $x$ the
transverse coordinate of a single particle in
the beam and by $p$ its conjugate momentum, we consider the following
single-particle hamiltonian
\bee
H = { p^2 \over 2} + U(x,z)~~~.
\label{2}
\ene
where $H$ has been made dimensionless dividing by the quantity
$m\gamma\beta^2c^2$ ($m$ and $\gamma$ being the particle rest mass and
the relativistic factor, respectively). In (\ref{2}) $U(x,z)$ stands for a
dimensionless potential filled by the particle.

\subsection{Main features of TWM}

As stated before \cite{fm1}-\cite{fmp},
in the {\it Thermal Wave Model} the collective behaviour
of a charged particle beam is ruled by a Schr\"{o}dinger-like
equation,
which can be obtained from the single-particle description by means of the
following {\it Thermal Quantization Rules} (TQR) \cite{fm1}-\cite{fm2},
\cite{fgm},
\bee
x \rightarrow \hat{x} ~~~, ~~p \rightarrow \hat{p} \equiv -i \epsilon
{ \partial \over \partial x}~~~,~~~
\mbox{and}~~H \rightarrow \hat{H} \equiv
i \epsilon {\partial \over \partial z}~~~.
\label{1}
\ene
where here $\epsilon$ stands for the transverse beam emittance.
Consequently, prescriptions (\ref{1}) give immediately
the following evolution
equation for the BWF
\bee
i \epsilon {\partial \over \partial z} \Psi(x,z) =
- { \epsilon^2 \over 2} { \partial^2 \over \partial x^2} \Psi(x,z)+
U(x,z) \Psi(x,z)~~~.
\label{4}
\ene
Provided that the normalization condition $\int_{-\infty}^{+\infty}
\left| \Psi(x,z)\right|^2~dx=1$ is satisfied,
the transverse number density of the beam particles $\Lambda (x,z)$ is given by
\bee
\Lambda (x,z) = N \left| \Psi(x,z)\right|^2~~~,
\label{5}
\ene
where $N$ is the total number of particles.
Eq.(\ref{4}) is formally identical to the paraxial
wave equation which describes the optical beam propagation in
inhomogeneous media with refractive index profile given by $U(x,z)$.
In this analogy the inverse of the wave number is replaced by the beam
emittance.

It has been shown in \cite{fm1} that in the simplest case of a relativistic
charged particle beam crossing a pure quadrupole (aberrationless lens)
of focusing strength $k_{1}>0$ ,
the BWF satisfies the following parabolic equation
\bee
i \epsilon {\partial \over \partial z} \Psi(x,z) =
- { \epsilon^2 \over 2} { \partial^2 \over \partial x^2} \Psi(x,z)+
{1 \over 2} k_{1}~x^2 \Psi(x,z)~~~.
\label{6}
\ene
It is easy to show that Eq. (\ref{6}) admits the following orthonormal
discrete modes
\begin{eqnarray}
\Psi_{n}(x,z)  =   \frac{1}{\left[2 \pi~\sigma^{2}(z)~ 2^{2n}~(n!)^2
\right]^{1/4}}~H_{n}\left( \frac{x}{\sqrt{2} \sigma(z)}\right)
\nonumber\\
 \times  \exp\left[ - \frac{x^2}{4 \sigma^{2}(z)} + i
\frac{x^2}{2 \epsilon \rho(z)} + i (1+2n) \phi(z) \right]~~~~\mbox{with}~~~
n=0,1,2,..~~~,
\label{13c}
\end{eqnarray}
where $H_{n}$ are the Hermite polynomials, $\sigma(z)$ obeys to the
following {\it envelope equation}
\bee
\sigma '' + k_{1}~\sigma + { \epsilon^3 \over 4
\sigma^2}=0~~~,
\label{14c}
\ene
and
\bee
\frac{1}{\rho}  = \frac{\sigma '}{\sigma}~~~,
\label{15ca}
\ene
\bee
\phi ' = -{ \epsilon \over 4 \sigma^2} ~~~.
\label{16c}
\ene
where each prime denotes the derivative with respect to $z$.
Within the similarity with electromagnetic beams, $\Psi_{n}(x,z)$
play the role analogous to the one played by Hermite-Gauss electromagnetic
modes, $\sigma(z)$ describes the {\it beam caustic} (i.e. {\it beam
envelope}), and $\rho(z)$ represent the
{\it wavefront curvature radius}. By introducing the r.m.s. $\sigma_{x}(z)$
associated to a general BWF satisfying (\ref{4}) as (the mean value of $x$
is assumed equal to zero)
\bee
\sigma_{x}(z) \equiv \langle x^2 \rangle^{1/2} = \left[\int_{-\infty}^{+\infty}
x^2 \left| \Psi(x,z)\right|^2~dx \right]^{1/2}
\label{17c}
\ene
(the quantum-like expectation value), it is easy to see that $\sigma(z)$,
appearing in (\ref{13c})-(\ref{16c}), coincides with the r.m.s. of
the fundamental mode $\Psi_{0}(x,z)$
\bee
\sigma(z) = \left[\int_{-\infty}^{+\infty}
x^2 \left| \Psi_{0}(x,z)\right|^2~dx \right]^{1/2}~~~.
\label{18c}
\ene
In addition, we can also define the expectation value for the transverse
linear momentum associated to $\Psi_{0}(x,z)$
\bee
\sigma_{p}(z) = \epsilon \left[\int_{-\infty}^{+\infty}
\left| { \partial \Psi_{0}(x,z) \over \partial x} \right|^2~dx
\right]^{1/2}~~~.
\label{19c}
\ene
By following particle accelerator physics language, $\sigma(z)$,
$\rho(z)$ and $\sigma_{p}(z)$ can be expressed in terms of some optical
parameters, called Twiss parameters $\alpha(z)$, $\beta(z)$, and $\gamma(z)$
\cite{twiss}
\bee
\sigma^2(z) = \epsilon ~\beta(z)~~~,
\label{20c}
\ene
\bee
{1 \over \rho(z)} = -{\alpha(z) \over \beta(z)}~~~,
\label{21c}
\ene
\bee
\sigma_{p}^2(z) = \epsilon~\gamma(z)~~~.
\label{22c}
\ene
It is suitable to introduce the following matrix
\bee
\hat{T}(z) \equiv \left( \begin{array}{cc} \gamma(z) & \alpha(z) \\
\alpha(z) & \beta(z) \end{array}\right)~~~.
\label{23c}
\ene
whose determinant, as
it is well-known \cite{twiss}, is constant with respect to z, namely
\bee
d\left[\mbox{det}\hat{T}(z)\right]/dz=0~~~.
\label{23d}
\ene
Consequently, we choose for our convenience, without loss of generality,
$\gamma \beta - \alpha^2=1/4$.
Consequently from (\ref{20c})-(\ref{22c}) follows that the
determinant of the matrix
\bee
\epsilon \hat{T}(z) \equiv \left( \begin{array}{cc} \sigma_{p}^2(z) &
- \sigma(z)~\sigma '(z) \\
- \sigma(z)~\sigma '(z) & \sigma^2(z) \end{array}\right)~~~.
\label{24c}
\ene
is an invariant, namely
\bee
\sigma_{p}^2 \sigma^2 - \left( \sigma \sigma '
\right)^2 = { \epsilon^2 \over 4} = \mbox{const.}~~~.
\label{25c}
\ene
It is easy to prove that
\bee
\sigma(z)\sigma '(z)= \int_{-\infty}^{+\infty}
\Psi_{0}^{*}(x,z) \left( { x \hat{p} + \hat{p} x \over 2}\right) \Psi_{0}(x,z)
{}~dx = \langle { x \hat{p} + \hat{p} x \over 2} \rangle ~~~.
\label{26c}
\ene
Consequently, from (\ref{25c}) follows that
\bee
\langle x^2 \rangle ~ \langle \hat{p}^2 \rangle -
\langle { x \hat{p} + \hat{p} x \over 2} \rangle^2 =
{\epsilon^2 \over 4}~~~,
\label{27c}
\ene
which is formally identical to Robertson-Schr\"{o}dinger uncertainty
relation \cite{rs1}, \cite{rs2} for partial cases when
$~\langle x\rangle =~\langle p\rangle =0$.
Furthermore, within the framework of TWM,
(\ref{27c}) is a quantum-like version of the well-known Courant-Snyder
invariant \cite{courant}. Note that (\ref{25c}), or equivalently (\ref{27c}),
gives immediately the usual form of the Heisenberg-like uncertainty principle
\cite{fm1} which is analogous to Heisenberg uncertainty relation in
quantum mechanics \cite{heis} (again for $~\langle x\rangle
=~\langle p\rangle =0$)
\bee
\sigma_{p} \sigma \geq { \epsilon \over 2}~~~.
\label{28c}
\ene
By introducing, for the general hamiltonian (\ref{2}), the total averaged
energy
associated to the transverse motion of the particles
\bee
{\cal  E}(z) \equiv \int_{-\infty}^{+\infty} \Psi^*(x,z) \hat{H}
\Psi(x,z)~dx =
{\epsilon^2 \over 2} \int_{-\infty}^{+\infty} \left| {\partial \Psi(x,z) \over
\partial x} \right|^2 ~dx + \int_{-\infty}^{+\infty} U~\left| \Psi(x,z)
\right|^2~dx~~~,
\label{29c}
\ene
the following virial equation
\bee
{d^2 \sigma_{x}^2(z) \over dz^2} = 4 {\cal E} - 2 \langle x {\partial U \over
\partial x} \rangle~~~,
\label{30c}
\ene
and the following energy-variation equation
\bee
{d {\cal E} \over dz} = \int_{-\infty}^{+\infty} { \partial U \over \partial z}
{}~\left| \Psi(x,z)\right|^2~dx
\label{31c}
\ene
hold. For a quadrupole-like potential (see Eq.(\ref{6})), Eqs.(\ref{30c}) and
(\ref{31c}) show that ${\cal E}$ is a constant of motion ($d {\cal E} /dz=0$),
and , in particular, for $\Psi = \Psi_{0}$ recover the envelope equation
(\ref{14c}). The equilibrium solution of (\ref{14c}) ($d^2
\sigma(z)/dz^2=0$), namely
\bee
\sigma_{0}^2 = { \epsilon \over 2 \sqrt{k_1} }~~~,
\label{32c}
\ene
plays an interesting role. In fact, (\ref{32c}) prescribes that if we
prepare the system, before entering the quadrupole-like device,
with given $\sigma_{0}$, $\epsilon$ and $k_{1}$ in such a way to
satisfy (\ref{32c}), the evolution of the beam through this device
is performed with the transverse size fixed at the equilibrium value
($\sigma(z)=\sigma_{0}$). In other words, (\ref{32c}) corresponds to
an initial beam configuration with the Twiss parameter $\alpha(z)=0$, i.e.
zero-divergence of the beam, or equivalently to a BWF with a
wavefront whose curvature radius is infinity. Thus, the beam
divergence (the curvature radius of the BWF wavefront) is also zero
(infinity) during the beam evolution.\\
Note that for a quadrupole-like potential and for the equilibrium
solution (\ref{32c}),  the set of Hermite-Gauss modes (\ref{13c})
reduces to the hamiltonian eigenstates of the harmonic oscillator
\bee
\Psi_{n}^{0}(x,z) = { 1 \over [ 2 \pi \sigma_{0}^2 2^{2n} (n!)^2]^{1/4}}
\exp\left( - { x^2 \over 4 \sigma^2_{0}} + i (1+2n) \phi_{0}(z) \right)~
H_{n}\left( { x \over \sqrt{2} \sigma_{0} }\right)~~~,
\label{7}
\ene
where $n=0,1,2,....$,
\bee
\phi_{0}(z)=- \sqrt{k_{1}} {z \over 2} ~~~,
\label{8}
\ene
and the energy values ${\cal E}_{n}^{0}$, given by (\ref{29c}) under the
substitution of these eigenstates, coincide with the hamiltonian
eigenvalues of the harmonic oscillator
\bee
{\cal E}_{n}^{0} = \left( n + { 1 \over 2} \right)\epsilon
\sqrt{k_{1}}~~~.
\label{8ab}
\ene
In particular, for $n=0$ (\ref{7}), (\ref{8}) and (\ref{8ab}) give
the ground-like state
\bee
\Psi_{0}^{0}(x,z) = { 1 \over [ 2 \pi \sigma_{0}^2]^{1/4}}
\exp\left( - { x^2 \over 4 \sigma^2_{0}} + i  \phi_{0}(z) \right)~~~,
\label{8a}
\ene
which is a pure real Gaussian and the lowest
energy reachable by the beam is ${\cal E}_{0}^{0} = (1/ 2)\epsilon
\sqrt{k_{1}}$. The means $\langle x\rangle $ and $\langle p\rangle $
are equal to zero at this state of beam.
In these conditions the uncertainty relation is
minimized as
\bee
\sigma_{0} ~ \sigma_{p0} = { \epsilon \over 2}~~~.
\label{8ac}
\ene
Eq. (\ref{8ac}) holds also during the evolution of the beam, because,
in addition to (\ref{32c}), we have $\sigma_{p}(z)=\sigma_{p0}=
\mbox{const.}$.
In summary, we conclude that if we initially prepare the beam
according to the matching conditions
(\ref{32c}), its evolution is ruled by a quantum-like behaviour in
terms of BWF-ground-like state which minimizes the uncertainty relation
and corresponds to the lowest accessible beam energy $(1/ 2)\epsilon
\sqrt{k_{1}}$.

By introducing the connection between emittance and transverse beam
temperature $T$, showed both in the conventional beam physics \cite{conv},
and in TWM \cite{fm1}, $\epsilon^2 /4\sigma_{0}^2 = k_{B} T/(m_{0} \gamma
\beta^2 c^2)$ ($k_{B}$ being the Boltzmann constant), we easily obtain
\bee
{\cal E}_{0}^{0} = ~{k_{B} T \over
m_{0} \gamma \beta^2 c^2}~~~,
\label{8ad}
\ene
where (\ref{32c}) has been used. Consequently, the beam energy
associated with the equilibrium solution corresponds to the thermal
energy and this result is in agreement with the equipartition theorem.

As we will show in subsection $2.3$,
BWF (\ref{8a}) belongs to the infinite series of coherent state functions,
labeled by a complex number $\alpha = \alpha_{1} + i \alpha_{2}$, and widely
used in quantum mechanics and  quantum optics \cite{gsk1}-\cite{gsk3}.
To further develop this point,
in the next subsection we present the general formalism of coherent
states and their main properties.

\subsection{Formalism of coherent states}

In fiber optics the complex ray formalism \cite{marc}, \cite{ar} has been
used to describe Gaussian wavepackets propagating along the fiber in frame of
Fock-Leontovich parabolic equation \cite{fl} which is a Schr\"{o}dinger-like
equation. In Refs. \cite{manko}, \cite{spring} it was proved that
the complex rays are
just the coherent states describing the electromagnetic beams in fibers.
Within Dirac's {\it bra} and {\it ket} formalism, the
coherent state $|\alpha \rangle$ is the state which represents the following
series
\bee
| \alpha \rangle = \exp\left( - { | \alpha|^2 \over 2} \right)
{}~\sum_{n=0}^{\infty} { \alpha^{n} \over \sqrt{n!}} ~| n \rangle~~~,
\label{8ae}
\ene
where the number state $|n \rangle$ corresponds to discrete modes of the
electromagnetic beam and $\alpha =\alpha _{1}+ i~\alpha _{2}$ is a
complex number. Since the following orthonormal condition holds
\bee
\langle n | m \rangle = \delta_{nm}~~~,
\label{8af}
\ene
we have the normalization condition for coherent states:
\bee
\langle \alpha | \alpha \rangle = 1~~~.
\label{8afbis}
\ene
A coherent state may be obtained using the unitary {\it shift
operator}
\bee
{\cal D}_\alpha = \exp\left(\alpha~ a^{\dag} - \alpha^{*}~a \right)~~~,
\label{8ag}
\ene
where the photon creation and annihilation operators $a^{\dag}$ and $a$
satisfy the boson commutation relation
\bee
\left[ a, a^{\dag}\right] =1~~~.
\label{8ah}
\ene
Thus, if one acts with this operator on the fundamental mode state
$|0\rangle$, such that
\bee
a~|0\rangle =0~~~,
\label{8ai}
\ene
we have the normalized coherent state
\bee
{\cal D}_\alpha | 0 \rangle = | \alpha \rangle~~~.
\label{8al}
\ene
Note that ${\cal D}_\alpha$ has the following shifting properties
\begin{eqnarray}
{\cal D}^{\dag}_\alpha ~a~{\cal D}_\alpha &  = & a + \alpha~~~,\label{8am}\\
{\cal D}^{\dag}_\alpha ~a^{\dag}~{\cal D}_\alpha &  = & a^* + \alpha^*
{}~~~.\label{8an}\\
\end{eqnarray}
{}From these formulas it follows, that the coherent state is the eigenstate of
the annihilation operator
\bee
a~|\alpha \rangle = \alpha | \alpha \rangle~~~.
\label{8ao}
\ene
If one calculates in the coherent state the means of quadrature components
$~Q$ and $~P$ one has
\bee
\langle\alpha |Q|\alpha \rangle =\frac {1}{\sqrt 2}
\langle \alpha |a + a^{\dag}|\alpha\rangle ={\sqrt 2}\alpha _{1}~~~,
\label{8ao1}
\ene
\bee
\langle\alpha |P|\alpha \rangle =\frac {\imath }{\sqrt 2}
\langle \alpha |a - a^{\dag}|\alpha \rangle ={\sqrt 2}\alpha _{2}~~~.
\label{8ao2}
\ene
The shift operator has the multiplication property
\bee
{\cal D}_\alpha ~{\cal D}_\beta  = {\cal D}_{\alpha+\beta}
{}~\exp\left(i Im[\alpha \beta^*] \right)~~~,
\label{8ap}
\ene
where $Im[...]$ stands for the imaginary part.
The distribution function of the electromagnetic modes
$|n \rangle$ in the
coherent state
\bee
P(n) = \left| \langle n | \alpha \rangle \right|^2 =
{\exp(-|\alpha|^2) \over n!} ~|\alpha |^{2n}~~~
\label{8aq}
\ene
is the Poisson distribution for which the expectation value of
mode number
\bee
\langle n \rangle = \sum_{n=0}^{\infty} n ~P(n) = |\alpha|^2~~~,
\label{8ar}
\ene
and the dispersion of the mode number
\bee
\sigma_{n} = \langle n^2 \rangle - \langle n \rangle^2 = |\alpha|^2
\label{8as}
\ene
is equal to the same value. For coherent states the dispersion of photon
quadrature components
\begin{eqnarray}
\langle\alpha|\left( { a + a^{\dag} \over \sqrt{2}} \right)^2 | \alpha \rangle
& -
\Bigr(\langle \alpha|
{a + a^{\dag} \over \sqrt{2}}  |\alpha\rangle \Bigr)^2 =
& { 1\over 2}~~~,\label{8at}\\
\langle\alpha|\left( { a - a^{\dag} \over i \sqrt{2}} \right)^2
| \alpha \rangle
& -
\Bigr(\langle \alpha|
{a - a^{\dag} \over i \sqrt{2}}
|\alpha\rangle \Bigr)^2 =
& { 1\over 2}~~~,\label{8au}
\end{eqnarray}
are equal and there is no correlation of these quadrature components
\bee
\langle\alpha | \left\{ a + a^{\dag}, a - a^{\dag} \right\} | \alpha \rangle
-2\langle\alpha |a + a^{\dag}|\alpha \rangle \langle\alpha |a
- a^{\dag}|\alpha \rangle =0
{}~~~,
\label{8av}
\ene
where $\{~,~\}$ denotes the anticommutator of two operators. In the coherent
state
representation the wave function of discrete mode state has the simple form
\bee
\langle \alpha | n \rangle = { \alpha^n \over \sqrt{n!}} \exp\left( -
{|\alpha|^2 \over 2} \right)~~~.
\label{8az}
\ene
As one sees from the decomposition (\ref{8ae}), the state $\exp(|\alpha|^2 /2)
| \alpha \rangle$ is the generating function for the number state $| n
\rangle$.
\\ According to (\ref{8at}) and  (\ref{8au}) the products of the
dispersions satisfies
the condition of minimizing the Heisenberg uncertainty relation. The
evolution
of the coherent state $|\alpha \rangle$ along the beam path due to the
hamiltonian of the oscillator $H = \omega~(a^{\dag} a + 1/2)$ produces
again the coherent state with extra phase factor
\bee
\exp\left[ - i \omega z \left( a^{\dag} a + { 1 \over 2} \right)\right]
{}~| \alpha \rangle = \exp\left[-i { \omega z \over 2} \right]
| \alpha \exp\left(-i \omega z  \right) \rangle~~~.
\label{8aaa}
\ene
{}From that point of view the coherent state is considered as the closest to
the classical state of the oscillator since it represents the wavepackets
whose center moves along the classical trajectory in the phase space and
the width of the packet is minimal and consistent with the uncertainty
relation \cite{gsk1}-\cite{gsk3}, \cite{mm}, \cite{dm}.

In the next subsection we will discuss the
properties of the coherent states in the coordinate representation
in order to describe the coherent charged particle beam transverse dynamics
within the framework of TWM.

\subsection{Physical meaning of coherent states for charged particle
beams}
One of the needs that we understand to introduce the notion of
coherent states for charged particle beams is connected with the
experimental possibility of producing, by acting with an external
electromagnetic force, a shifting off-axis of the beam propagation
coordinate. For example, in an accelerating machine this can be done
by means of some devices, such as
kickers, RF cavities, etc. \cite{bryant}-\cite{turner}, which
are able to produce a displacement of the transverse beam distribution
center from its stationary position (design orbit or synchronous
particle orbit) to another neighbouring path. This means that the
center of transverse beam space-distribution has a shift and this
effect allows us to naturally introduce, within the framework of TWM,
the coherent state representation.In fact, let us suppose that the beam is in
the ground-like state $\Psi_{0}^{0}(x,z)$ given by (\ref{7}) for $n=0$,
which describes the equilibrium state of the particle beam travelling
along the {\it stationary orbit} \cite{sands}.
In order to take into account the
displacement of the transverse distribution center we introduce the
following modified potential well
\bee
U(x,z)~=~{1 \over 2}k_{1}x^2~-~F_{1}(z)x~+~F_{2}(z)~~~,
\label{100}
\ene
where $F_{1}(z)$ and $F_{2}(z)$ account for the external
electromagnetic forces. Since $U(x,z)$ is still quadratic with respect
to $x$, from general properties of Schr\"{o}dinger equation, the
initial ground-like state $\Psi_{0}^{0}$ continues to be Gaussian, but its
center (wavepacket center) is now shifted off-axis.
In order to explicitely find this transformed BWF, let us consider
the following transformations on (\ref{4}) with $U$ given by
(\ref{100}):
\bee
\Psi
(x,z)=\exp\left[-{i\over\epsilon}\int_{0}^{z}F_{2}(z')~dz'\right]\Theta
(x,z)
\label{101}
\ene
which allows us to get
\bee
i \epsilon {\partial \over \partial z} \Theta(x,z) =
- { \epsilon^2 \over 2} { \partial^2 \over \partial x^2} \Theta(x,z)+
\left({1 \over 2} k_{1}~x^2 - F_{1}(z)~x\right)\Theta(x,z)~~~.
\label{102}
\ene
(note that $|\Psi (x,z)|=|\Theta (x,z)|$).
It is very easy to prove that (\ref{102}) has the following Gaussian solution
\bee
\Theta_{0}(x,z)~=~\left[{1\over 2~\pi~\sigma^{2}_{0}}\right]^{1/4}
\exp\left[-{\left(x-x_{0}(z)
\right)^{2}\over4~\sigma^{2}_{0}}~+~{i\over\epsilon}p_{0}(z)~x~
-i\delta_{0}(z)\right]
\label{102c}
\ene
where $\sigma_{0}$ still satisfies the equilibrium condition
(\ref{32c}) and
the functions $x_{0}(z)$, $p_{0}(z)$, and $\delta_{0}(z)$
satisfy the following differential equations
\bee
x_{0}''~+~k_{1}x_{0}~=~F_{1}(z)~~~,
\label{106a}
\ene
\bee
p_{0}''~+~k_{1}p_{0}~=~F_{1}'(z)~~~.
\label{106b}
\ene
and
\bee
\delta_{0}'~=~{p_{0}^2\over 2\epsilon}~-~\sqrt{k_{1}} {x_{0}^{2}\over
4\sigma_{0}^{2}}~+~{\sqrt{k_{1}}\over 2}~~~.
\label{106c}
\ene
Some considerations are in order:\\
i). Eqs.(\ref{106a}) and (\ref{106b}) coincide with the constraints
for the following canonical
transformation $x,p\rightarrow \tilde{x},\tilde{p}$, with
\bee
\tilde{x}=x-x_{0}(z)
\label{103}
\ene
\bee
\tilde{p}=p-p_{0}(z)
\label{104}
\ene
which transforms the hamiltonian $H=p^2+{1 \over 2} k_{1}~x^2 -
F_{1}(z)~x$ into $\tilde{H}=\tilde{p}^2+{1 \over 2}
k_{1}~\tilde{x}^2$.\\
ii). Consequently, $x_{0}$ and $p_{0}$
represent the space-coordinate
and the momentum-coordinate shift, respectively. For the following it
is suitable to introduce the following {\it complex dimensionless
shift}
\bee
\alpha (z)~\equiv {x_{0}(z)\over
2\sigma_{0}}~+~i{\sigma_{0}p_{0}(z)\over \epsilon}
\equiv \alpha_{1} (z)+i\alpha_{2} (z)~~~.
\label{108}
\ene
iii). Eq.(\ref{106a}) (Eq.(\ref{106b}))
can be thought as classical motion
equation of unit mass particle which feels the restoring force
$-k_{1}x_{0}$ ($-k_{1}p_{0}$) and the external force
$F_{1}(z)$ ($F_{1}'(z)$). \\
iv). It is very easy to see that, by virtue of (\ref{32c}), solution
(\ref{102c}) still minimizes the uncertainty relation, and,
consequently, describes a coherent state associated to the shifts
$x_{0}$ and $p_{0}$.\\
v). According to the definition (\ref{5}), solution (\ref{102c})
gives the following coherent-state-particle-distribution
$\Lambda_{0} (x,z)$
\bee
\Lambda_{0}
(x,z)~=~N~|\Theta_{0}(x,z)|^2~=~{N\over\sqrt{2\pi\sigma_{0}^{2}}}
\exp\left[-{\left(x
-2\sigma _{0}\alpha_{1}(z)\right)^2\over 2\sigma_{0}^{2}} \right] ~~~.
\label{110}
\ene

Thus, (\ref{102c}) can be cast in the following form
\bee
\Theta_{0}(x,z)~=~\left[{1\over 2~\pi~\sigma^{2}_{0}}\right]^{1/4}
\exp\left[{-x^{2}
\over 4~\sigma^{2}_{0}}\right]~
\exp\left[{\alpha (z)~x~\over \sigma_{0}}~-~{|\alpha (z)|^2\over 2}~-~
{\alpha^2 (z)\over 2}\right]~\exp\left[i\theta (z)\right]~~~,
\label{111}
\ene
where
\bee
\theta (z)~\equiv~\int_{0}^{z}\left[{\sigma_{0}\over
\epsilon}\alpha_{1}(z')F_{1}(z')~-~{\sqrt{k_{1}}\over 2}\right]~dz'~~~.
\label{112}
\ene
Note that, according to the formalism used in subsection $2.2$, the BWF
$\Theta_{0} (x,z)$
is a coherent state {\it produced} by the complex shift $\alpha (z)$.
In the Dirac's formalism, let us both denote this coherent state
as $|\alpha\rangle$, and introduce the notation
\bee
\langle x | \alpha\rangle \equiv\Theta_{0}(x,z)~~~,
\label{8b}
\ene
For the sake of simplicity
we will consider now these states at $z=0$, namely
\bee
\langle x | \alpha_{0}\rangle
\equiv\Theta_{0}(x,0)\equiv\Phi_{\alpha_{0}}(x)=
\left[{1\over 2\pi\sigma^{2}_{0}}\right]^{1/4}
\exp\left[{-x^{2}
\over 4\sigma^{2}_{0}}\right]
\exp\left[{\alpha_{0} ~x\over \sigma_{0}}-{|\alpha_{0} |^2\over 2}-
{\alpha_{0}^{2} \over 2}\right]~~~,
\label{8bb}
\ene
where $\alpha_{0}\equiv\alpha (0)\equiv\alpha_{1}(0)+i\alpha_{2}(0)
\equiv\alpha_{10}+i\alpha_{20}$.
These functions have the following properties \cite{gsk1}-\cite{gsk3}
\bee
\langle \alpha_{0} | \beta_{0} \rangle = \int^{+ \infty}_{-\infty}
\Phi^{*}_{\alpha_{0}}(x) \Phi_{\beta_{0}}(x) ~dx= \exp\left(
- { |\alpha_{0}|^2 \over 2}- { |\beta_{0}|^2 \over 2}+ \alpha_{0}^{*}
\beta_{0}\right)~~~,
\label{8d}
\ene
which means nonorthogonality, and
\bee
{ 1 \over \pi} \int_{-\infty}^{+\infty}d\alpha_{10}
\int_{-\infty}^{+\infty} d\alpha_{20}~| \alpha_{0} \rangle \langle
\alpha_{0} | =
\hat{1}~~~,
\label{8da}
\ene
which implies the completeness condition, i.e. any state $| \Phi \rangle$
with the wave function $\Phi(x)$ may be represented as a superposition of the
coherent states:
\bee
|\Phi \rangle = { 1 \over \pi} \int_{-\infty}^{+\infty}d\alpha_{10}
\int_{-\infty}^{+\infty} d\alpha_{20}~\langle \alpha_{0}|\Phi\rangle ~
|\alpha_{0}\rangle~~~,
\label{8e}
\ene
where the BWF in coherent state representation $\langle \alpha_{0}|\Phi
\rangle$
is given by the overlap integral
\bee
\langle \alpha_{0} | \Phi \rangle =
\int_{-\infty}^{+\infty}\Phi^{*}_{\alpha_{0}}(x) \Phi(x)~dx~~~.
\label{8f}
\ene
Since the overlap integral
of two different coherent states is not equal to zero
these states form an overcomplete set of functions. The particle beam
coherent state
$\Phi_{\alpha_{0}}(x)$ is a normalized eigenstate of the annihilation
operator
\bee
\hat{a} = { 1 \over \sqrt{2}} \left({ x \over \sqrt{2}\sigma_{0}} +
\sqrt{2}\sigma_{0}{\partial \over \partial x} \right)~~~,
\label{8g}
\ene
i.e.
\bee
\hat{a}~\Phi_{\alpha_{0}}(x) = \alpha_{0}~\Phi_{\alpha_{0}}(x)~~~.
\label{8h}
\ene
Since TWM is described for small
intensities by the linear Schr\"{o}dinger-like equation, any BWF associated
to a charged particle beam can be represented as
a continuous
superposition of coherent states, or discrete superposition
of Fock mode states.

\section{Charged particle beam dynamics through an infinite
1-D quadrupole with small sextupole and octupole deviations}

In this section we describe the transverse dynamics of the charge particle
beam while it is travelling through an infinite 1-D quadrupole with
small {\it sextupole} and {\it octupole} deviations ({\it
aberrations}). We consider here the same kinematic and geometrical
assumptions made in the introduction of section $2$.

Without aberrations the beam is assumed to satisfy the matching
condition (\ref{32c}) (unperturbed beam). Thus, according to the
results of section $2$, if the initial ($z=0$) transverse beam
distribution is purely Gaussian, i.e.
\bee
\Psi_{0}^{0}(x,0) = { 1 \over [ 2 \pi \sigma_{0}^2]^{1/4}}
\exp\left( - { x^2 \over 4 \sigma^2_{0}} \right)~~~,
\label{113}
\ene
the unperturbed beam evolution at any $z>0$ is described by a
particular coherent state which is the ground-like state (\ref{8a}) that we
can obtain from (\ref{102c}) for the special case $x_{0}=p_{0}=0$ or,
equivalently, from (\ref{8bb}) for $\alpha_{0}=0$ :
\bee
\Psi_{0}^{0}(x,z) = { 1 \over [ 2 \pi \sigma_{0}^2]^{1/4}}
\exp\left( - { x^2 \over 4 \sigma^2_{0}}\right)\exp\left(-{i\over
2}\sqrt{k_{1}z}\right)~~~.
\label{114}
\ene
In order to take into account small sextupole and octupole
aberrations, we perturb the (\ref{6}) into the following equation:
\bee
i \epsilon {\partial \over \partial z} \Psi(x,z) =
- { \epsilon^2 \over 2} { \partial^2 \over \partial x^2} \Psi(x,z)+
{1 \over 2} k_{1}~x^2 \Psi(x,z)+ \hat{V}~\Psi(x,z)~~~,
\label{9}
\ene
with the initial condition (\ref{113}),
where the perturbation
\bee
\hat{V}(x)= { 1 \over 3!}~k_{2}~x^3 + { 1 \over 4!}~k_{3}~x^4~~~,
\label{10}
\ene
accounts for the aberrations. In particular ${ 1 \over 3!}~k_{2}~x^3$
($ { 1 \over 4!}~k_{3}~x^4$) corresponds to the 1-D sextupole
(octupole) potential term. Note that $\hat{V}(x)$ given by (\ref{10})
can be considered a small perturbation provided that the following
conditions
\bee
k_{2}\sigma_{0}/3k_{1}<<1~~~{\mbox{and}}~~~k_{3}\sigma_{0}^{2}/12k_{1}<<1
\label{115}
\ene
hold.
In the next subsection we will give an approximate solution of the
problem (\ref{9}) with the specifications (\ref{113}), (\ref{10}), and
(\ref{115}).

\subsection{Analytical results: perturbative approach}

By using the standard perturbation approach we write the following expansion
of the $\Psi(x,z)$ in terms of the eigenstates $\Psi_{m}^{0}$
\bee
\Psi(x,z) = \sum_{m=0}^{+\infty} c_{m}(z)~\Psi_{m}^{0}(x,z)~~~.
\label{11}
\ene
The substitution of Eq. (\ref{11}) into (\ref{9}) yields the infinite set
of equations
\bee
i \epsilon { d c_{n} \over d z} = \sum_{m=0}^{+\infty} c_{m}(z)~
\langle n | \hat{V} | m \rangle~~~,
\label{12}
\ene
where $| n \rangle \equiv \Psi_{n}^0$. To solve the set of equation
(\ref{12}) we consider the first order correction to the BWF in the
case in which $c_{m}(0)=\delta_{m,0}$, according to the initial
condition (\ref{113}). If we write $ c_{m}(z)=
\delta_{m,0}+ c_{m}^{1}(z)$ (conditions (\ref{115}) here correspond to
the condition $|c_{m}^{1}|<<1$), Eqs. (\ref{12}) give
\bee
i \epsilon { d  \over d z} c_{n}^{1}(z) = \langle n | \hat{V} | 0 \rangle~~~,
\label{13}
\ene
where
\bee
\langle n | \hat{V} | 0 \rangle \equiv \int_{-\infty}^{+\infty}
\Psi_{n}^{0*}(x,z) \hat{V}(x) \Psi_{0}^{0}(x,z)~dx~~~.
\label{14}
\ene
Hence, the coefficients $c_{n}^{1}(z)$ are given by
\begin{eqnarray}
c_{0}^{1}(z) & = & - i{ 3\over 4}\mu\xi~~~,
\label{15a}\\
c_{1}^{1}(z) & = & - i{3 \over 2\sqrt{2}}\nu
{}~\exp\left[i {\xi \over 2} \right]
{}~\sin\left( {\xi \over 2}\right)~~~,
\label{15b}\\
c_{2}^{1}(z) & = & - i{ 3\over 4}\mu
{}~\exp\left[i\xi\right]
{}~\sin\left( \xi\right)~~~,
\label{15c}\\
c_{3}^{1}(z) & = & - i{1\over 12\sqrt{2}}\nu
{}~\exp\left[i {3\xi \over 2} \right]
{}~\sin\left( {3\xi \over 2}\right)~~~,
\label{15d}\\
c_{4}^{1}(z) & = & - i{ 1\over 32}\mu
{}~\exp\left[i2\xi\right]
{}~\sin\left(2 \xi\right)~~~,
\label{15e}
\end{eqnarray}
where we have introduced the dimensionless parameters
$\nu\equiv k_{2}\sigma_{0}/3k_{1}$ and
$\mu\equiv k_{3}\sigma_{0}^{2}/12k_{1}$,
and the dimensionless length $\xi\equiv \sqrt{k_{1}}z$.
The non-normalized perturbed BWF is then obtained by the superposition
(\ref{11}) for four terms only:
\begin{eqnarray}
\Psi(x,z) & = & \left[ ( 1 + c_{0}^1 (z))\Psi_{0}^{0}(x,z)+
c_{1}^1 (z) \Psi_{1}^{0} (x,z) + c_{2}^1 (z) \Psi_{2}^{0} (x,z) \right.
\nonumber\\
& + & \left. c_{3}^1 (z) \Psi_{3}^{0} (x,z) +c_{4}^1 (z) \Psi_{4}^{0} (x,z)
\right]~~~.
\label{16}
\end{eqnarray}
Note that now the BWF $\Psi (x,z)$ given by (\ref{16}) is not Gaussian
anymore and,
consequently, does not represent a coherent state anymore. In fact,
the aberrations introduce a defocusing of the particles which produces
a distortion of the particle beam distribution with respect the
unperturbed Gaussian profile.

In the next subsection we analyze more accurately this distortion on
the basis of numerical results.

\subsection{Numerical results}

According to the perturbation theory results obtained in the previous
subsection, the parameters $\mu$ and $\nu$ represent a measure of the
distortion due to sextupole and octupole, respectively, once the multipole
distortion is defined as the ratio of the sextupole (octupole) aberration
strength to the quadrupole strength. Provided that $\mu<<1$ and $\nu<<1$, the
distribution in configuration space is proportional to $|\Psi|^2$, where $\Psi$
is given by (\ref{16}). In $Fig.1$ we have plotted the normalized transverse
density profile versus the dimensionless transverse coordinate
$x/\sqrt{2}\sigma_{0}$ for increasing values of $\xi$ and for $\mu=.005$ and
$\nu=.05$ (see $Figs.1a-1f$). The dashed lines give the starting distribution.
We can clearly see a weak distortion of the particle distribution as the beam
propagates throughout the optical device, which is quite evident in $Figs.1b,
1d$, and $1f$: the solid lines represent the distorted distributions for
$\xi=15.5, 21.5, 27.5$, respectively. As $\xi$ increases monotonically, the
distortion increases and decreases alternatively. In particular in $Figs.1a,
1c$, and $1e$ ($\xi=12.5, 18.5, 24.5$, respectively), the distortion is
negligeable (negligible distortion), while $Figs. 1b, 1d$, and $1f$ show
significative distortion. This effect is due to the fact that, according to
(\ref{16}), the BWF is given by a superposition of five modes which interfere
each other. To clarify this point in $Fig.2$ we have plotted the evolution of
the normalized distortion of the transverse density profile as a function of
the dimensionless longitudinal coordinate $\xi$. The normalized distortion is
defined here by $\Delta \sigma/\sigma_{0}$, where
$\Delta\sigma\equiv\sigma(\xi)-\sigma_{0}$ is the deviation of the effective
transverse beam size (r.m.s.) evaluated at a generic $\xi$ from the initial one
$\sigma_{0}$. In $Fig.2$ we have fixed $\nu=.05$ and plotted the distortion for
$\mu=0, .0015, .002$. When $\mu=0$, the octupole gives no contribution and the
BWF is a superposition of three modes only (see Eq.(\ref{16})). From $Fig.2$ is
quite evident that the oscillating behaviour of the distortion increases if we
increase $\mu$.\\ In $Fig.3$ we have plotted the evolution of the distortion
fixing $\mu=.005$ and for $\nu=0, .05 $. When $\nu=0$, we have no contribution
from the sextupole, but, when both sextupole and octupole deviation are taken
into account, we clearly see the enhancement of the distortion of the beam
profile.

\section{Remarks and conclusions}

In this paper we have introduced the concept of {\it coherent state}
for charged particle beams within the framework of TWM
\cite{fm1}-\cite{fmp}.

We have shown that the coherent Gaussian-like structures for charged
particle beams,
naturally or artificially generated in particle accelerators
\cite{bryant}-\cite{turner},
correspond to examples of beam wave functions which have a
formal definition identical to those ones are widely used in the
quantum mechanics and quantum optics \cite{gsk1}, \cite{gsk3}.

We have used the analogy among {\it quantum mechanics}, {\it electromagnetic
beam optics in optical fibers}, and
{\it particle beam dynamics} in the framework of TWM in order
to introduce the concept and to understand
the physical meaning of coherent state for a charged particle
beam. We have shown that such a kind of state is associated with the complex
parameter $\alpha$, introduced in section $2$, which describes the shift
of the beam distribution center from the optical axis. In particular, we
have proved that when the charged particles of a purely Gaussian beam,
travelling in an accelerating machine, are
shifted from their path by the action of some external transverse
electromagnetic forces \cite{bryant}-\cite{turner},
the state after this action is just
the coherent state $|\alpha\rangle$
(the real part of $\alpha$ corresponds to the shift of the space
coordinate $x$ and the imaginary part of $\alpha$ corresponds to the shift of
the conjugate momentum $p$).

Physically, due to the action of an external electromagnetic force,
the distribution center results to be shifted.
This shift, in turn, transforms the initial BWF modulus
$|\Psi_{0}^{0}(x,z)|$ into
the following: $|\Psi (x,z)|$=$|\Psi_{0}^{0}(x-x_{0},z)|$.
In addition, since the shifting does not change the equilibrium
condition (\ref{32c}), because it is an intrinsic relationship
between the quantities $\sigma_{0}$, $\epsilon$, and $k_{1}$, and
since the BWF profile is still Gaussian, the uncertainty relation
initially fixed at its minimum value $\epsilon$ is preserved during
this shift.
But this picture, in terms of formalism and physics of charged
particle beams, fully corresponds to the coherent state description
given in subsection $2.2$, and the transformed BWF describes a new
coherent state with respect to the simplest one which corresponds to
the ground-like state.
{}From the experimental point of view the shift operator ${\cal D}_\alpha$
(\ref{8ag}), which produces these new coherent states for
charged particle beam, in an accelerator can be realized
by means of some devices such as kickers, RF
cavities, etc. \cite{bryant}-\cite{turner}.

Furthermore, by taking the ground-like-state as the initial condition, we have
studied the particle beam evolution through a infinite 1-D
quadrupole-like in the presence of small sextupole and octupole
deviations. We have shown that, in the framework of TWM,
this evolution is governed by a 1-D Schr\"{o}dinger-like equation for
the BWF which have been solved by means of the standard
time-dependent perturbation techniques. To the first-order, this
perturbative treatment has shown that, during the evolution, the
particle beam profile does not correspond to a coherent state anymore.
This is due to the aberrations which produce a distortion of the
initial Gaussian profile: during the beam evolution through the
optical device the BWF is a superposition of four Hermite-Gauss
eigenstates only. The interference
of these four states produces, during the evolution, {\it small
oscillating distortions} of the particle beam profile around the
ground-like state. Numerical estimates for this effect have also been
given.

In a forthcoming paper
we will consider the present perturbative treatment
by taking, instead of the ground-like state ($\alpha_{0}=0$),
a general coherent state ($\alpha_{0}\neq 0$) as initial
condition. In addition, since in quantum optics there were introduced other
quantum states which are generalization of coherent ones, such as squeezed
and correlated states \cite{hal}-\cite{sudar},
we will discuss their analogs in particle beam physics.

\newpage

{\Large \bf Figure captions}

\bigskip

\begin{itemize}
\item[Fig.1]~Normalized transverse profile of the beam density
$\sqrt{2\pi}\sigma_{0}|\Psi|^2$ vs. $x/\sqrt{2}\sigma_{0}$ for
increasing value of $\xi$ and for $\nu=.05$ and $\mu=.005$.\\
Dashed line~=~starting distribution, solid line~=~distorted
distribution\\
(a) $\xi=12.5$; (b) $\xi=15.5$; (c) $\xi=18.5$; (d) $\xi=21.5$; (e) $\xi=24.5$;
(f) $\xi=27.5$;

\item[Fig.2]~Normalized distortion $\Delta\sigma /\sigma_{0}$ vs. $\xi$ for
fixed $\nu=.05$ and $\mu=0., .0015, .002$.

\item[Fig.3]~Normalized distortion $\Delta\sigma /\sigma_{0}$ vs. $\xi$ for
fixed $\mu=.005$ and $\nu=0., .05$.

\end{itemize}

\end{document}